# Studying on Opinion Evolution by Hamilton-Jacobi Equation

Chen-Jie Feng[1], Peng Wang[1,2], Jie Huo[1,3], Rui Hao[1,3] and Xu-Ming Wang[1,3,*]

1 School of Physics and Electrical Information Engineering, Ningxia University, Yinchuan, 750021, PR China
2 Department of General Studies, Beifang University of Nationalities, Yinchuan, 750021, PR China
3 Ningxia Key Laboratory of Intelligent Sensing for Desert Information, Yinchuan, 750021, PR China
* E-mail: wang_xm@126.com

## Abstract

A physical description of an opinion evolution is conducted based on the Hamilton-Jacobi equation derived from a generalized potential and the corresponding Langevin equation. The investigation mainly focuses on the heterogeneities such as age, connection circle and overall quality of the participants involved in the opinion exchange process. The evolutionary patterns of opinion can be described by solution of the Hamilton-Jacobi equation, information entropy. The results show that the overall qualities of the participants play critical roles in forming an opinion. The higher the overall quality is, the easier the consensus can reach. The solution also demonstrates that the age and the connection circle of the agents play equally important roles in forming an opinion. The essence of the age, overall quality, and connection circle corresponds to the maturity of thought (opinion inertia), reason and intelligence, influence strength of the environment, respectively. So the information entropy distributes in the phase planes (two variable combinations out of the three---age, connection circle and overall quality) exhibit the evolution rules of an opinion.

## 1. Introduction

Human behaviors, such as decision-making and opinion formation, are extremely complex; because they are often determined by interactions between multiple aspects of the participants, such as personal experience, inference and will. This is one reason why the dynamic origin of an opinion evolution is difficult to reveal. From another point of view, it is that much attention of scientists has been drawn. So there have been many attempts to investigate the manner in which an opinion evolves.

The most valuable attempts have been conducted by physicists to investigate the opinion formations from multiple perspectives. The so-called voter models owe the change of an individual opinion to the influence of his/her neighbors [1-4]. The Ising model, which is used to model ferromagnetism with ferromagnetic spins in solid-state physics, is suggested to describe the evolution of binary opinions [5-8]. To describe the situation in which a agent only takes the agents into account if the opinion differences between his and the agents' are confined with a certain range (a confident level), the bounded confidence model is proposed [9, 10]. To simulate voting process under the majority principle, the majority rule model is suggested. The main rule is that an agent is inclined to follow local viewpoints [11]. Obviously, the aforementioned models demonstrate analogous spirits, that is, the opinion of an agent updates following that of his neighbors. So the simulations show the same result, consensus. Later on, many investigations on opinion formation are performed via optimizing and modifying the classic models mentioned above [12-18]. Another typical situation is disagreement or opinion differentiation. In order to realistically reflect such characteristics caused by some complicated coupling between the agents and the environment, Shao introduced the memory effect into the majority rule model, and reproduced the co-existence of the binary opinions with unbalanced weights [19].

Schweitzer describes the collective opinion formation and the spatial migration with the so-called active Brownian particles via placing the agents in a communication field [20]. The analysis results present some social characteristics that can be understood by the analogies in physical systems. This may be the biggest advantage of such a general physical investigation. Similar studies have been conducted by Helbing [21, 22], Martins [23, 24] and Wang [25]. They devoted to looking for some general descriptions of the opinion evolution under some general physical principles.

In this article, we introduce a generalized potential (GP) to express the communication field in which an agent exchanges his opinion with the others, i.e., the environment. The GP is defined by the age, connection circle and overall quality of the agent. The collective opinion evolution can be described by solution of the Hamilton-Jacobi (H-J) equation which is derived from the GP. The solution presents the dependence of the information entropy on the aforementioned three factors. We hoped thereby to understand the collective opinion evolution. This method actually follows an optimal way in which a natural system evolves generally due to the fact that the H-J equation originally comes from Euler-Lagrange equation restricted by the principle of least action. The method was also applied to investigating the group migration behavior [25] and protein folding [26, 27].

## 2. From Generalized Potential to Hamilton-Jacobi Equation

The GP that can embody the underlying mechanism of interactions among individuals highlights the heterogeneity of individuals such as the age, connection circle and overall quality. These ingredients are related directly to the opinion evolution. 1) Age: It denotes the opinion inertia of an agent. As the age increases, one may hold steadier opinion since he can see a thing more objectively and rationally. One may say that the older one is the more mature and perfect his value-judgment system will be, and then he is able to keep more opinion away from change. 2) Connection circle: It represents the influencing strength of the environment acting on the agents. For a given agent, it may be positively correlated with the number of the participants. As the connection circle extends, one will be apt to strengthen his opinion or change his opinion due to the increased persuaders with the opposite altitudes, support or opposition. 3) Overall quality: It embodies the cognitive ability of an agent, which may be associated to the personal education, regional economy, local culture, etc. Lower overall quality may lead one to lack of information and lower cognitive ability. Conversely, higher overall quality can make one to make appropriate judgments.

Based on the discussions above, the GP function can be defined as

$$U(r,a,b) = c_1 e^r + c_2 \ln(1+a) + c_3 b^{-\alpha} + c_{12} e^r \ln(1+a) + c_{13} e^r b^{-\alpha} + c_{23} \ln(1+a) b^{-\alpha}. \quad (1)$$

Where $a, r$ and $b$ denotes mean age of the participants, connection circle of an agent and the overall quality of the participants, respectively. $c_1, c_2, c_3$ are the coupling factors. $c_1, c_2$ are positive and $c_3$ is negative. $c_{12}, c_{13}, c_{23}$ are the crossing coupling strength between the three factors.

In order to analyze opinion evolution, we introduce Langevin equation to describe the dynamics of opinion. It reads

$$\frac{\partial x}{\partial t} = -2D \nabla U(x) + \eta(t) \;, \quad (2)$$

where $x = (r, a, b)$ is a state vector, $U(x)$ is actually the GP expressed by Equ. (1), $D$ is the diffusion coefficient and variable $\eta(t)$ denotes the Gaussian white noise that is characterized by mean value $\langle \eta(t) \rangle = 0$. And the second-order correlation function

$$\langle \eta(t)\eta(t') \rangle = 2D\delta(t-t') \quad (3)$$

As well known, the stochastic dynamic equation (1) can be easily transformed into the Fokker-Planck Equation,

$$\frac{\partial p}{\partial t} = 2D \frac{\partial}{\partial x}\left(\frac{\partial U}{\partial x} p\right) + D \frac{\partial^2 p}{\partial x^2} \;, \quad (4)$$

Where $p$ is transition probability of the opinion change from initial state $x_i$ at $t_i$ to the final state $x_f$ at $t_f$. In other words, the solution of Fokker-Planck equation is subjected to the boundary conditions $x_i$ and $x_f$, and can be written in terms of a path-integral,

$$p(x_f, t_f | x_i, t_i) = e^{-\frac{U(x_f) - U(x_i)}{2}} \int_{t_i}^{t_f} Dx(\tau) e^{-S_{eff}[x]} d\tau \quad (5)$$

where

$$S_{eff}[x] = \int_{t_i}^{t_f} \left[ \frac{1}{2}\left(\frac{\partial x}{\partial t}\right)^2 + \frac{D^2}{2}\left(\frac{\partial U}{\partial x}\right)^2 - D^2 \frac{\partial^2 U}{\partial x^2} \right] d\tau, \tag{6}$$

and it is called the effective action. We set

$$V_{eff}(x) = \frac{D^2}{2}\left(\frac{\partial U}{\partial x}\right)^2 - D^2 \frac{\partial^2 U}{\partial x^2}, \tag{7}$$

and it is named after the effective potential. This quantity often measures the tendency of a configuration to evolve under Langevin diffusion. In fact, the probability for the system to remain in the same configuration $x$ under an infinitesimal time interval is given by

$$p(x, d\tau | x, 0) = e^{-V_{eff}(x) d\tau} \tag{8}$$

Obviously, the states of high effective potential are extremely unstable in the Langevin diffusion process. Applying the normalization condition, Lagrangian $\int dx p(x,t|x_i,t_i) = 1$, we have

$$1 = \int_{-\infty}^{+\infty} p(x_f, t_f | x_i, t_i) dx_f = e^{-U(x_i)/2} \int_{-\infty}^{+\infty} \int_{x_i}^{x_f} e^{\frac{U(x_f)}{2} - \frac{S_{eff}[x]}{D}} dx(\tau) dx_f. \tag{9}$$

According to the integral principle of steepest descent, one may seek out, the most probable integral path for Equ. (8) is the one that satisfies the Euler-Lagrange equation. So we can obtain the effective Lagrangian function, $L_{eff} = \frac{\dot{x}^2}{2} - V_{eff}$, with proper boundary condition

$$\ddot{x} = 2 \frac{\partial V_{eff}}{\partial x}. \tag{9}$$

The solution of the dynamic equations represents a point of the evolution with time in the phase space. In fact, we could think it is a canonical transformation from a point $x(t)$ to the fixed point $x(t_i)$ in the phase space. Here, the integral of upper limit $t_f$ in the Equ.(6) is turned into the arbitrary time $t$, and different time denotes different orbits of evolution. It is called the main function of Hamilton and read as,

$$S = \int_{t_i}^{t} L_{eff}(x_c, \dot{x}_c, t) dt. \tag{10}$$

It reflects a variety of possible paths, and has a characteristic of entropy.

Human behaviors could analogy to the classical particle system. So, it is satisfied with the distribution of Boltzmann. It is written as,

$$p(x_f, t_f | x_i, t_i) = \frac{e^{-U(x_f)/K_B T}}{Z}, \tag{11}$$

where $Z$ is a partition function.

$$Z = \int e^{-\frac{U(x)}{K_B T}} \theta \times (K_B T - (U(x) - U(x_c))) dx. \tag{12}$$

Using the information entropy $S_{inf} = -\frac{\partial}{\partial T} K_B T \ln Z$, we obtain

$$L_{eff} \propto S_{in} \tag{13}$$

Detailed derivation could be referred to the Ref. [25]. Thus, the main function $S$ is called the information entropy. Then, the operator in the Equ.(10) is made the variation, read as,

$$S = \int_{t_i}^{t_f} \left( \frac{\delta L_{eff}}{\delta x} \delta x + \frac{\delta L_{eff}}{\delta \dot{x}} \delta \dot{x} \right) dt. \tag{14}$$

According to the intrinsic properties of differential operator is written as,

$$\frac{dS}{dt} = \frac{\partial S}{\partial t} + \frac{\partial S}{\partial x} \dot{x} = L_{eff} \tag{15}$$

And

$$L_{eff} - \frac{\partial S}{\partial x} \dot{x} = -H \tag{16}$$

This equation together with Equ.(15) implies

$$\frac{\partial S}{\partial t} = -H\left(x, \frac{\partial S}{\partial x}, t\right) \tag{17}$$

In principle, a solution of the Hamilton-Jacobi equation obtains by the multi-dimensional steepest descent method.

$$\frac{\partial S}{\partial t} = -TrH\left(x, \frac{\partial S}{\partial x}, t\right) \tag{18}$$

Where $TrH\left(x, \frac{\partial S}{\partial x}, t\right)$ is the trace of Hessian matrix based on the function of

Hamilton $H = \frac{1}{2}\dot{x}^2 + U$.   Equ. (18) is now written as

$$\frac{\partial s}{\partial t} = \frac{1}{(1+a)^2}(c_2 + c_{12}e^r + c_{23}b^{-\alpha}) - e^r(c_1 + c_{12}\ln(1+a) + c_{13}b^{-\alpha})$$

$$-\alpha(\alpha+1)b^{-\alpha-2}(c_3 + c_{13}e^r + c_{23}b^{-\alpha})$$

$$\frac{\partial^2 r}{\partial t^2} = 2D^2 e^r(c_1 + c_{12}\ln(1+a) + c_{13}b^{-\alpha}) \cdot (e^r(c_1 + c_{12}\ln(1+a) + c_{13}b^{-\alpha}) - 1) \quad (19)$$

$$\frac{\partial^2 a}{\partial t^2} = 2D^2(-\frac{1}{(1+a)^3})(c_2 + c_{12}e^r + c_{23}b^{-\alpha}) \cdot (c_2 + c_{12}e^r + c_{23}b^{-\alpha} + 2)$$

$$\frac{\partial^2 b}{\partial t^2} = 2D^2 \alpha(\alpha+1)b^{-\alpha-3}(c_3 + c_{13}e^r + c_{23}\ln(1+a)) \cdot (-\alpha b^{-\alpha}(c_3 + c_{13}e^r + c_{23}\ln(1+a)) + \alpha + 2)$$

## 3. Simulation results and equivalents in opinion evolution

Using 4th-order Runge-Kutta method to solve the above system of equations, we can obtain the information entropy varying with age, connection circle, overall quality and time. It is well-known that the information entropy can characterize the opinion dynamics by actually tracking the evolution group structure formed by the opinion exchanges among the participants. The following presented solutions hold for a large scale of the combinations of the parameters. Here we only present the results when the values of parameters are chosen as $c_1 = 1.2 \times 10^{-7}$, $c_2 = 1.1 \times 10^{-5}$, $c_3 = 1.2 \times 10^{-5}$, $c_{12} = 3.7 \times 10^{-6}$, $c_{23} = 2.9 \times 10^{-5}$, $c_{13} = -1.5 \times 10^{-5}$, $\alpha = 1.2 \times 10^{-5}$.

To comprehensively demonstrate the dependence of information entropy on age, connection circle and overall quality, we prefer phase diagram to show the details of the relationship.

As shown in Fig.1, the dependence of information entropy on age and connection circle is depicted by the contours in the age-connection circle plane. It is evident that the plane can be divided into two regions. In the first region, the left-upper part, the information entropy possesses small values and varies with age in a small range. This indicates that in a small connection circle, consensus can be easily formed in a small connection circle due to the fact that each agent is exerted relative small influences from the others. In the direction of age increasing, information entropy decreases and then increases. This implies that the consensus cannot be easily formed due to the fact that opinion changes easily and frequently for the young agents; the consensus can be formed relatively easily since the adults have accumulated a series of methods or set up rational judgments and therefor express their opinions clearly; the consensus can be also not easily formed for the elderly agents because they often have rigid minds and changeless doctrines.

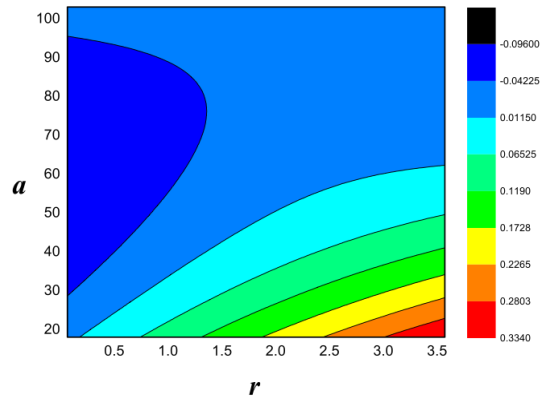

Figure 1. Information entropy in age-connection circle phase plane.

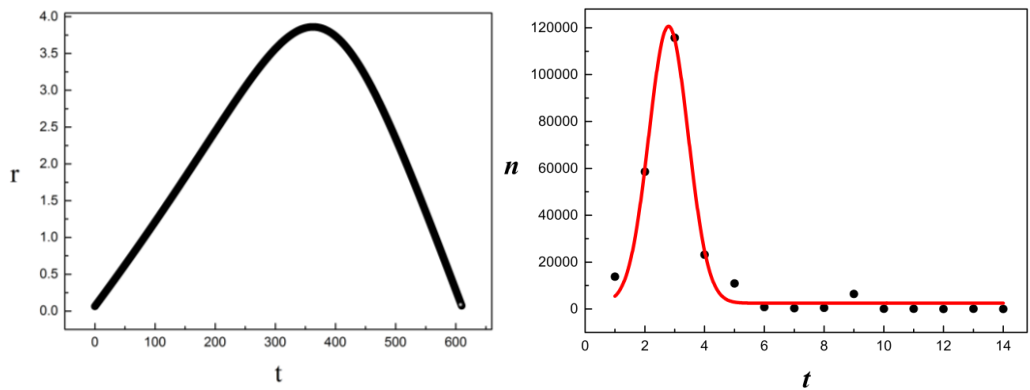

Figure 2. The variation of the connection circle with time (left); the number of the Weibo comment about the events of the woman-driver in Chengdu varies with time (right).

In the second region, the right-lower part, the information entropy mainly depends on the connection circle, and monotonously increases in the direction of connection circle extending. And the gradient in this direction follows a single peak since the color intervals firstly increased and then decreased, which indicates that the opinions become more and more difficult to reach consensus as the connection circle extends since the environment meanwhile gets more and more complicated for each of the participants.

It is interesting that the variation of the connection circle scale with time follows a single peak (Fig. 2 (left)). This maybe correspond to a real world situation, that is, the number of the participants will firstly increase and then decrease for an open social circle in an opinion exchange process. In other words, some event, message or viewpoint can usually draw more and more attentions and even to culminate, and then go out of focus. Such a correspondence exists everywhere, for instance, the number of the micro-blog comments on the conflict caused by the behavior that a woman driver in Chengdu frequently changed lanes in May 3, 2015 (Fig. 2 (right)). This correspondence also means that our attempt can grasp some essential aspects of opinion evolution.

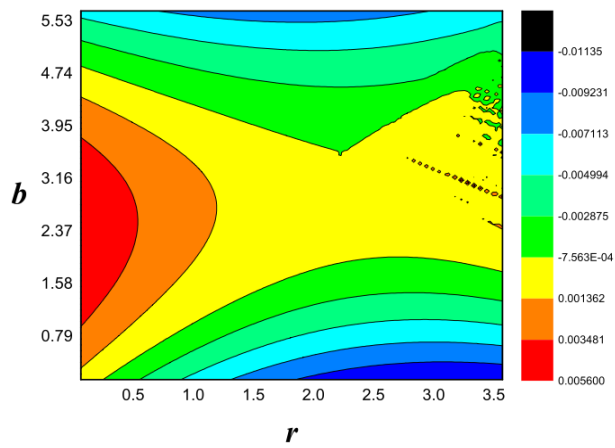

Figure.3 Information entropy in overall quality-connection circle phase plane.

Fig. 3 shows the information entropy patterns formed by the contours in $b-r$ plane. In the direction of the overall quality increasing, the information entropy depicts a single peak of large value in a small connection circle. It is not difficult to understand that the agents possessing high overall quality will insist on their own mature points of view, so the opinions of each agent are not easily changed. The high value of information entropy embodies such a disorder status. As the overall quality is great than some threshold, the perfect judgment of the agents can reduce this irregularity. When the connection circle extends, this irregularity is also slightly reduced due to the increase of the environment influence. And the complicated environment of extended connection circle may lead to the patterns more complex, the islands on the right-upper region are the representatives of such complexity. However, the feature of the single peak remains. In the direction of overall quality improving, the phase space can be divided into three parts. The first part locates in $b \in [1.8, 4.2]$, in which and in the direction of connection circle extending, information entropy has high value due to the fact that everyone, with high overall quality, holds his own point of view, so that it is very difficult to form consensus. Therefore, irregularity and stability is the main feature of the region. As connection circle extends, the increased influence of the environment can depress this irregularity, and the value of information entropy has the value decreased. The second part is that $b < 1.8$, and the third part is $b > 4.2$ and is below the leading diagonal. In these two regions and the direction of connection circle extending, information entropy firstly decreases and then increase, which indicates that for a given overall quality a moderate size of connection circle is better for forming an opinion. A small size of the connection circle may lack the leading point of view, while a big size may cause the influence of the environment more complex. The consensus can be slightly difficult to form. It is also interesting that in a larger range of connection circle, the curve in information entropy-overall quality space takes on a single peak. This implies that for a given size of the connection circle, the agents of lower overall quality often show the so-called group psychology, if there is a leading point of view, a consensus can be easily formed; the agents of higher overall quality often possess objective and rational judgment, and the therefore hold similar points of view, a consensus can be, of course, easily formed. In contrast, the medium overall quality of the agents makes the opinion evolution process demonstrating "a hundred schools of thought contend" due to their uneven judgments to a thing. So the consensus is difficult to form.

At last we present the results in $b - a$ phase plane. The prominent feature is the patterns formed by contours at the two corners, the left-upper one and the right-lower one. These structures correspond to the monotonous increase of information entropy in the direction of age increasing at a small value of overall quality and the same variation trend of information entropy in the direction of overall quality strengthening at a small value of age. This similarity indicates that the increases of both age and overall quality represent the reason and intelligence which can be easily transformed into pig-headedness. At the two corners mentioned above, this situation reaches the extreme, the opinion is difficult to form. However, there is a valley along the leading diagonal in which the contributions to the information entropy from age and overall quality are equivalent. So this region has small value of information entropy, and the "cooperation" of the age and overall quality make a consensus more easily formed.

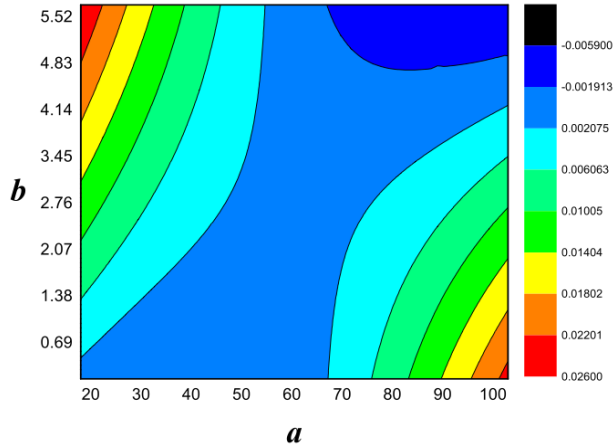

Figure 4. The information entropy distributed in the personal background-age plane

## 4. Conclusions

In summary, we propose a physical method, from generalized potential to the H-J equation, to investigate collective opinion evolution by tracking the variation of the information entropy. The dependences of information entropy on the age, overall quality and connection circle of an agent are depicted by contours in phase plane. The results present some new understandings to opinion evolution characteristics and the corresponding mechanisms. The biggest contribution may lies in the uncovering the essences of the age, connection circle and overall quality of the agent in the opinion evolution process. Age denotes the opinion inertia of an agent; connection circle represents the strength of influence of the environment on the agents; overall quality embodies the cognitive ability of an agent. So the phase planes, $a-r$, $b-r$ and $b-a$ demonstrate many patterns of opinion evolution. By the aid of these patterns, we obtain the physical descriptions and deep understandings of the collective opinion evolution via interpreting the relationship between the information entropy and the three fundamental variables, age, connection circle and overall quality. It is worth pointing out that way in which the number of the participants varies with time is supported by the empirical statistical result about the Weibo comment number changes in a similar manner. So this attempt may provide us with a reference role to investigating other related systems.

**Acknowledgment:** This study is supported by National Natural Science Foundation of China under Grant No. 11265011.